 \documentclass[aps,prl,preprint,groupedaddress]{revtex4-1}
\usepackage{graphicx}  
\usepackage{dcolumn}   
\usepackage{bm}        
\usepackage{amssymb}   
\usepackage{amsmath}
\usepackage{xcolor}

\begin{document}

\title{Nonlinear transmission of laser light through coronal plasma due to self-induced incoherence}
\author{A. V. Maximov, J. G. Shaw, J. P. Palastro} 
\affiliation{
Laboratory for Laser Energetics, University of Rochester, 
Rochester, New York 14623, USA
}
\date{\today}

\begin{abstract}

The success of direct laser-driven inertial confinement fusion (ICF) relies critically on the efficient
coupling of laser light to plasma. At ignition scale, the absolute stimulated Raman scattering (SRS)
instability can severely inhibit this coupling by redirecting and strongly depleting laser light. This Letter 
describes a new dynamic saturation regime of the absolute SRS instability. The saturation occurs
when spatiotemporal fluctuations in the ion-acoustic density detune the instability resonance. The
dynamic saturation mitigates the strong depletion of laser light and enhances its transmission through the
instability region, explaining the coupling of laser light to ICF targets at higher plasma densities.

\vspace{0.2cm}
\end{abstract}

\maketitle

In direct-drive inertial confinement fusion (ICF), an ensemble of laser beams symmetrically illuminates a
cryogenic target containing thermonuclear fuel \cite{SC}. The illumination ionizes and heats the outer shell 
of the target, creating a pressure that drives inward fuel compression and outward mass ejection. 
The mass ejection creates a region of low-density plasma, or corona, that plays a critical role in direct-drive ICF: coupling of 
laser light to the corona determines the strength of the ablation pressure and, ultimately, the implosion 
performance \cite{SC,VG}. 

A high ablation pressure requires the transmission of laser light to deep within the corona where collisions 
can efficiently convert electromagnetic energy to plasma thermal energy. To get there, however, the laser light must 
propagate through the outer corona, where it can drive a number of scattering processes that grow as parametric 
instabilities \cite{KR,JM}. In their nonlinear stage, these instabilities can redirect the incident light into unwanted 
directions and repartition the light energy into plasma modes. These modes can undergo local collisional damping, in which 
case the energy is deposited too far from the ablation surface, or collisionless damping, which creates  nonthermal 
electrons that can preheat the fuel and reduce its compressibility \cite{SC,VG1}. In either case, the premature depletion of laser 
energy in this region presents a significant challenge for direct-drive ICF. 

Among the laser plasma instabilities, stimulated Raman scattering (SRS) has the lowest intensity threshold for a broad 
range of ICF conditions \cite{KR,DL,A}, making it a commonly observed feature in both direct- and indirect-drive ICF experiments
\cite{JK,DS1,M,MR}. In SRS, an incident laser light wave decays into a scattered Raman light wave and an electron plasma wave. 
In a region near the quarter of the critical plasma density $n_c = m_e \omega_0^2 / 4 \pi e^2$, determined by the frequency of the laser $\omega_0$, 
where $m_e$ and $e$ are the electron mass and charge, respectively, 
the SRS decay waves can grow exponentially in time as an absolute instability until they nonlinearly saturate \cite{DL,A}. 
Depending on the plasma parameters, this saturation has been predicted to take several forms, such as the Langmuir decay instability (LDI)  
(when an electron plasma wave decays into another electron plasma wave and an ion-acoustic wave) \cite{DG,HK,KRT,SD,JP1}, 
particle trapping \cite{VU1,VU,Y1,DS,CEA}, or SRS rescattering \cite{BW}. 

Recent planar-target experiments on the National Ignition Facility (NIF) that emulated the plasma corona of an ignition-scale direct-drive 
implosion showed a clear SRS feature originating from close to the quarter-critical density \cite{MR}. The observations confirmed theoretical 
estimates that, because of the large density scale lengths in the plasma corona ($\sim$few hundred microns), the threshold for absolute SRS 
would be exceeded. Those same estimates also suggest that the instability would strongly deplete the laser light, preventing 
significant transmission deep into the corona. The hydrodynamic evolution of the target was consistent, however, with the efficient conversion of 
laser energy into plasma thermal energy \cite{MR}. As a result, a critical question emerges: How can the laser light propagate through the 
absolute instability region with a high transmission rate?

This Letter describes, for the first time, a dynamic saturation regime of the absolute SRS instability 
due to self-induced incoherence.
As the incident light propagates through the instability region, 
it drives a primary SRS decay that initially depletes the laser intensity. 
The electron plasma waves resulting from the SRS decay then undergo a secondary instability
that drives a broad spectrum of low-frequency density perturbations. 
The instability saturates when the density perturbations reach a high enough level 
to detune the primary SRS resonance, establishing
a dynamic balance between the transmitted and scattered laser light. 
This dynamic, incoherent saturation mitigates depletion and facilitates the transmission of the laser light 
through the instability region, explaining how light can penetrate deep into the corona to efficiently drive ICF implosions. 

The saturation of absolute SRS was investigated using the laser-plasma simulation environment ({\it LPSE}) \cite{LP,LPSE,JP2}. 
{\it LPSE} employs a fluid plasma model to 
describe the evolution of the four waves (light, Raman, electron plasma, and ion-acoustic) and the couplings
between them. The 
simulations were performed in two spatial dimensions ($x$ and $y$) with s-polarized light
(the electric field vectors of both the pump and Raman light waves were perpendicular to the simulation plane). The laser light was normally 
incident on a plasma with linear gradient $n=\frac{1}{4}n_{c}(1+x/L)$, where $L$ is the scale length at the quarter-critical density. The 
parameters, listed in Table \ref{table:t1}, correspond to plasma conditions relevant to NIF experiments. 

\begin{table}[]
\begin{tabular}{l | l }
\hline \hline
Laser wavelength ($\lambda_0$)     & 0.351 $\mu m$  \\
\hline
Scale length ($L$)          & 500 $\mu m$    \\
\hline
Electron temperature ($T_e$)                  & 4 keV          \\
\hline
Ion temperature ($T_i$)                  & 4 keV \\
\hline
Ion charge ($Z$)                    & 3.5 \\
\hline
Ion atomic number ($A$)                   & 2 $Z$ \\                  
\hline
Density range        & $(0.21 \ to \ 0.265) \ n_c$ \\
\hline \hline                                    
\end{tabular}
\caption{Ignition-relevant NIF parameters used in the {\it LPSE} simulations.}
\label{table:t1}
\end{table}

\begin{figure}[ht!]
\centering
\includegraphics[width=3.6in]{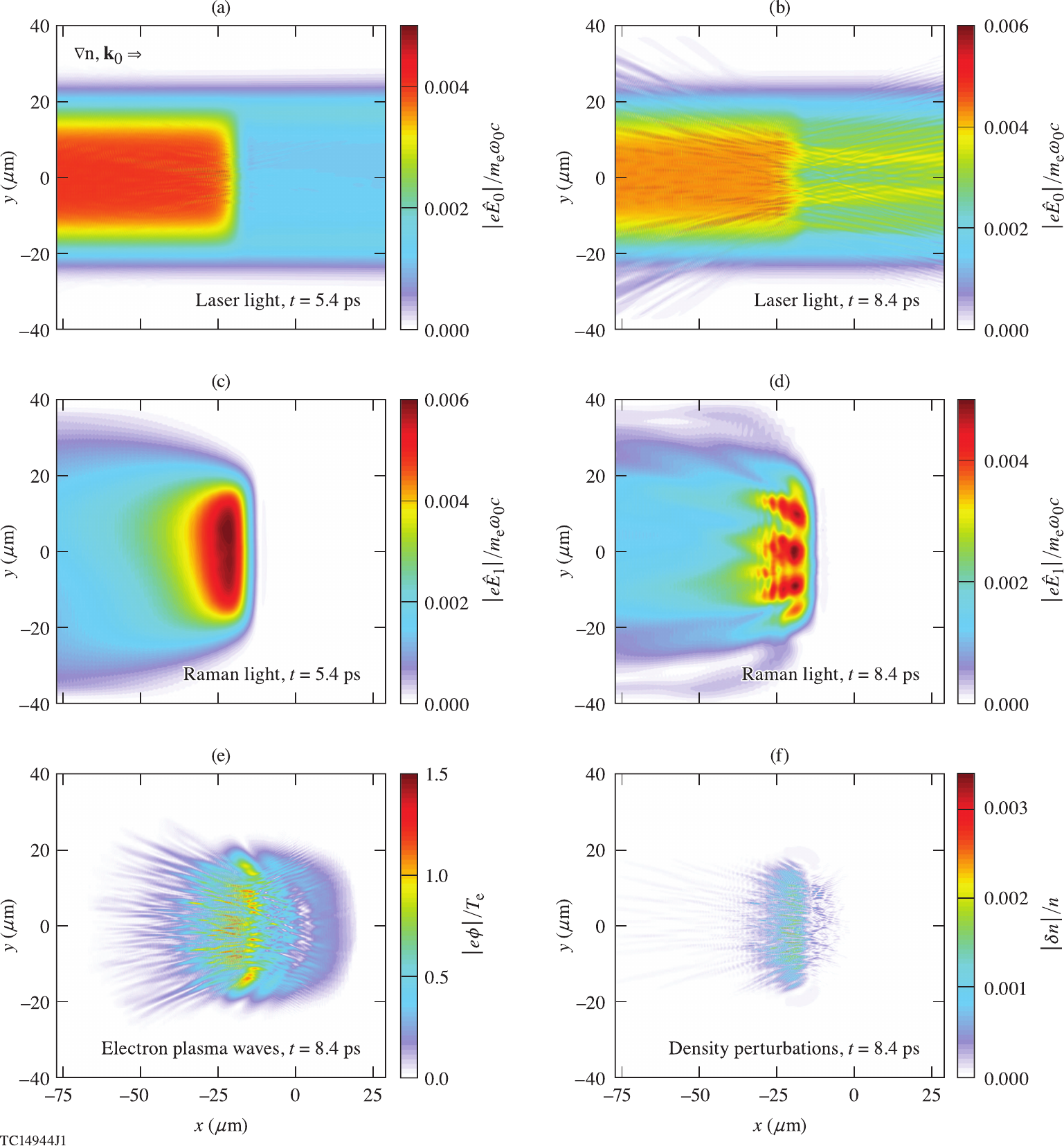}
\caption{ The amplitudes of laser and Raman light waves at 5.4 ps
((a) and (c) respectively) and 8.4 ps ((b) and (d) respectively).
The amplitudes of plasma waves (e) (in terms of wave potential energy normalized to temperature) 
and low-frequency density perturbations (f) at 8.4 ps.}
\label{fig:f1}
\end{figure}

Figures \ref{fig:f1}(a) - \ref{fig:f1}(d) illustrate the nonlinear saturation of absolute SRS for laser light with an incident intensity of
$I_0 = 2 \times 10^{14} \ W/cm^2$ (approximately $3 \times$ greater than the theoretical instability threshold at
these parameters, $7 \times 10^{13} \ W/cm^2$ \cite{DL}).
Early in time, the laser light propagates through the plasma without scattering. Shortly thereafter, the absolute instability develops. 
By 5.4 ps,  the instability has strongly depleted the pump [Fig. \ref{fig:f1} (a)], and the Raman light [Fig. \ref{fig:f1} (c)]
has grown to an amplitude comparable to the laser light. This pump depletion stage quickly gives way, 3 ps later, to a dynamic saturation stage in which the amplitudes of both the laser and Raman light 
become nonstationary and spatially incoherent [Figs. \ref{fig:f1}(b) and \ref{fig:f1}(d), respectively].  

\begin{figure}[ht!]
\centering
\includegraphics[width=1.8in]{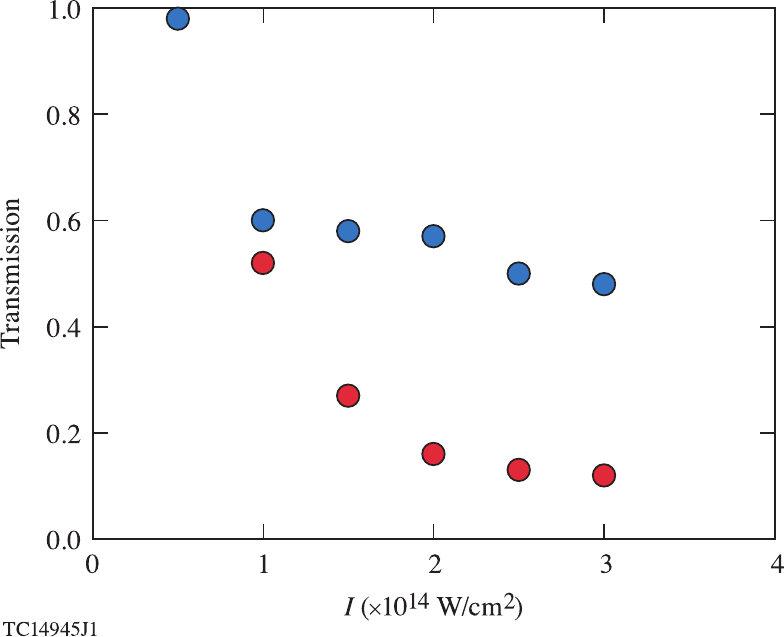}
\caption{The scaling of light transmission at the end of the pump depletion stage (red points) and during
the dynamic saturation stage
(blue points) as a function of the incident laser intensity.}
\label{fig:f2}
\end{figure}

The induced spatial incoherence increases the laser light transmission $T$ from $T = 0.16$ 
in the pump depletion stage to $T = 0.59$ in the dynamic saturation stage---
a near-$4 \times$ increase in power transported deeper into the plasma corona. 
Figure \ref{fig:f2} displays the scaling of the transmission, in both the pump depletion (red points) 
and dynamic saturation (blue points) stages, as a function of laser intensity. 
At intensities below the absolute SRS threshold ($I_0 < 7 \times 10^{13} \ W/cm^2$), 
the transmission is reduced by only a small factor due to collisional absorption. 
Above the threshold, the dynamic saturation increases the transmission well above the levels determined 
by pump depletion alone. Reflection of the laser light accounts for only $\sim 5\%$ of the drop in transmission. 

The development of spatial incoherence in the laser light can be understood as a two-step process: 
First, the electron plasma waves amplified during absolute SRS [Fig. \ref{fig:f1} (e)] undergo a transverse scattering instability, 
which ponderomotively drives low-frequency ion fluctuations [Fig. \ref{fig:f1} (f)]. 
Second, the resulting low-frequency fluctuations seed near-forward stimulated Brillouin scattering (SBS) of the incident light wave, 
i.e., the decay of a light wave into a scattered light wave and ion-acoustic wave. 

\begin{figure}[ht!]
\centering
\includegraphics[width=3.4in]{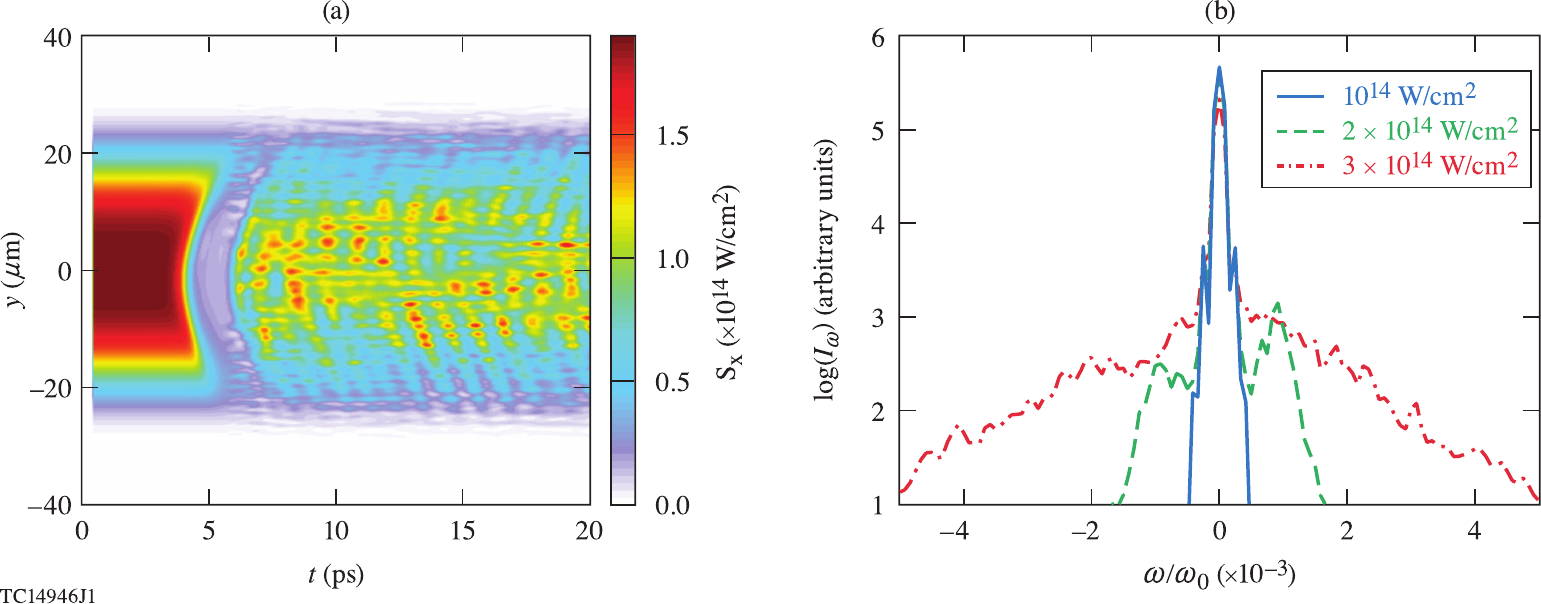}
\caption{Transmitted Poynting flux as a function of time and transverse coordinate $y$ 
for the laser intensity $2 \times 10^{14} \  W/cm^2$ (a), 
and the frequency spectra of transmitted light for three incident laser intensities 
of $10^{14} \ W/cm^2$ (blue line), $2 \times 10^{14} \  W/cm^2$ (green line), 
and $3 \times 10^{14} \  W/cm^2$ (red line) (b).}
\label{fig:f3}
\end{figure}

The transmitted Poynting flux [Fig. \ref{fig:f3}(a)] and light spectra [Fig. \ref{fig:f3}(b)] 
demonstrate the presence of near-forward SBS. 
Figure \ref{fig:f3}(a) shows the longitudinal component of the transmitted Poynting flux 
exiting the simulation region as a function of time and transverse coordinate $y$ 
for $I_0 = 2 \times 10^{14} \ W/cm^2$. 
During the dynamic saturation stage (time $>$ 7 ps), Poynting flux exhibits spatial incoherence with a scale of $ \ell \sim 2 \lambda_0$, where
$\lambda_0 = 2\pi c/\omega_0$, and temporal oscillations with a period $\tau \sim 1 \textnormal{ps}$. The two are connected by the ion-acoustic sound speed 
$c_s \sim \ell/\tau $, illustrating that the low-frequency ion-acoustic fluctuations spatiotemporally modulate the laser light. 
Here $c_s = \sqrt{(Z T _e + 3 T_i)/ m_i}$, where $Z$ is the ion charge number, $T_e$ and $T_i$ are the electron and ion temperatures, respectively,
and $m_i$ is the ion mass. 

Figure \ref{fig:f3}(b) displays the frequency spectra of the transmitted light calculated 
for three different incident intensities during the saturation stage (time $>$ 7 ps). 
Consistent with the increase in transmission, the temporal incoherence, 
observable as the width of the spectrum, increases with intensity. At the lowest intensity (blue line), 
$I_0 = 10^{14} \ W/cm^2$, the spectrum is narrowly peaked about the incident frequency. 
Stokes and anti-Stokes features appear at an intensity of $2 \times 10^{14} \ W/cm^2$ (green line). 
These sidebands have a spectral shift of  $\approx 10^{-3} \omega_0$, corresponding to the 1-ps oscillations 
in the Poynting flux and the presence of near-forward SBS. 
At the largest intensity (red line), $I_0 = 3 \times 10^{14} \ W/cm^2$, the spectrum has significantly broadened: 
the large amplitude of the SRS-generated plasma waves drives the transverse scattering instability to a highly nonlinear state, 
creating a broad spectrum of ion-acoustic fluctuations that enhance the incoherence. 

\begin{figure}[ht!]
\centering
\includegraphics[width=3.6in]{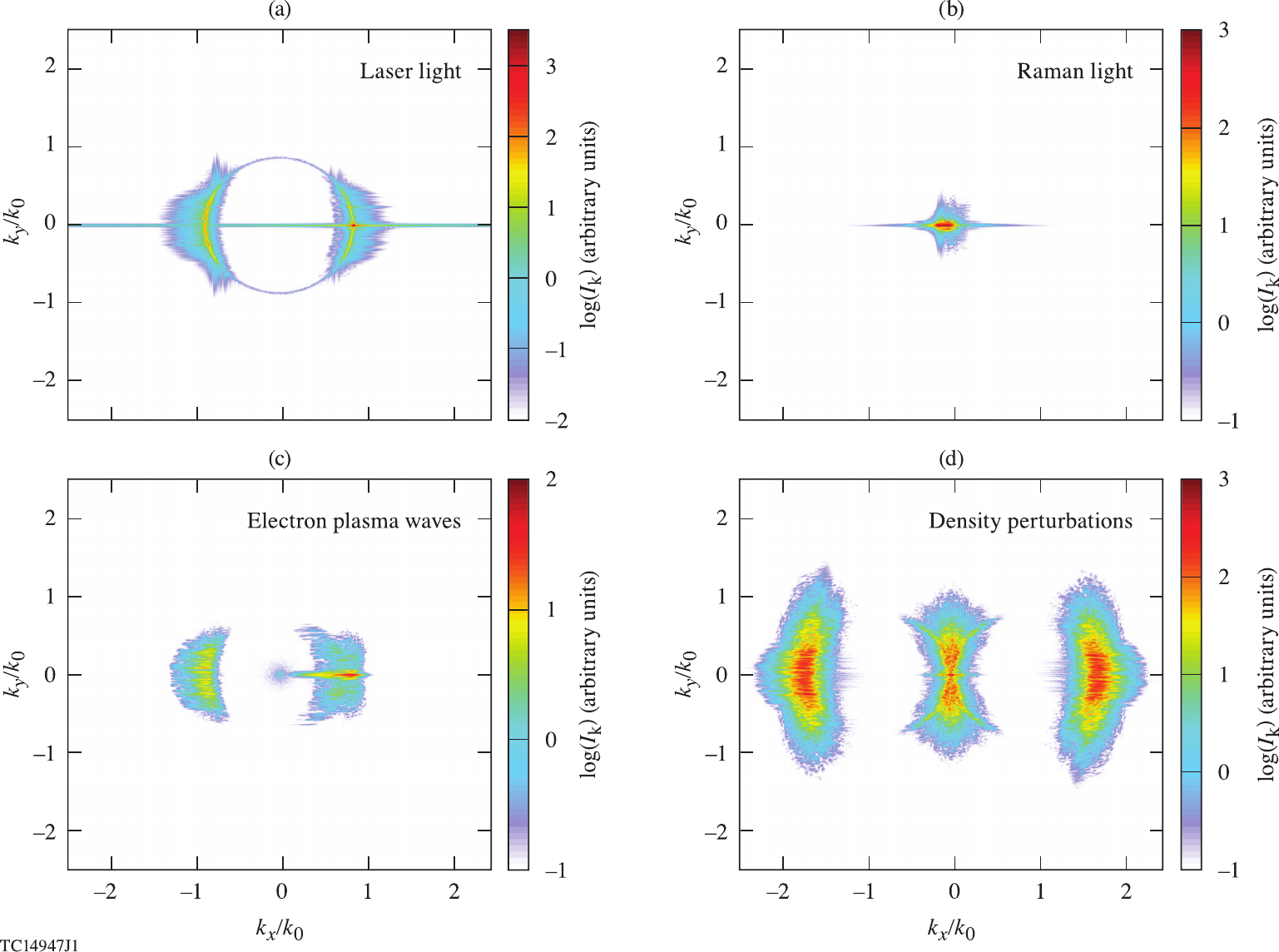}
\caption{
The spectra of laser light waves (a), Raman light waves (b),
plasma waves (c) and low-frequency density perturbations (d) in the dynamic saturation stage (at 8.4 ps) for the parameters of Fig. \ref{fig:f1}.}
\label{fig:f4}
\end{figure}

Distinct features of the instabilities are clearly observed in the spatial spectra of the waves,
illustrating the primary SRS decay, the transverse instability of the electron plasma waves, and near-forward SBS [Figs. \ref{fig:f4}(a)-\ref{fig:f4}((d)].
These figures display the spectra for the laser light, Raman light, electron plasma waves, and low-frequency density fluctuations,
respectively.
Foremost, the primary decay products of absolute SRS can be identified as the bright features near zero $k$ for the Raman light 
[Fig. \ref{fig:f4}(b)] and near $k_x \sim k_0 $ and $k_y = 0$ for the electron plasma wave [Fig. \ref{fig:f4}(c)]. 
Here $k_0 = 2\pi /\lambda_0$ is the wave vector of the incident laser in vacuum.

The electron plasma waves generated from SRS undergo a decay instability in which their ponderomotive force drives ion-acoustic fluctuations 
in the direction transverse to the density gradient. 
The fluctuations then small-angle scatter the electron plasma waves, which, in turn, reinforces the transverse ponderomotive force. 
These phenomena can be observed in Figs. \ref{fig:f4}(c) and \ref{fig:f4}(d) as spreading in $k_y$ 
about $k_x \approx k_0$ for the electron plasma waves and about $k_x \approx 0$ for the ion-acoustic waves. 
Figures \ref{fig:f1}(e) and \ref{fig:f1}(f) show the associated transverse ripples in the electrostatic potential of the electron plasma waves 
and the density of ion-acoustic waves, respectively. Both waves are co-localized with the SRS instability region.

The simulations described by Figs. \ref{fig:f1}-\ref{fig:f4} were performed in the plasma region of 116 by 80 $\mu m$, 
with the incident beam width of 40 $\mu m$. Simulations with a wider plasma and incident laser beam exhibited similar instability evolution and 
dynamic saturation.

The dispersion relation for the transverse plasma wave instability (with ion-acoustic wave vector $k_y$) 
can be obtained from the equations for the secondary plasma waves and the ion-acoustic perturbations \cite{DG}. 
At its resonance, the dispersion equation for the transverse instability has the form:
\begin{equation}
\left( \gamma + \gamma_{i}  \right) \cdot \left( \gamma + \gamma_p  \right) 
= {\omega_e \over 4} (k_{y} c_s) I_p
\label{eq5},
\end{equation}
where $\gamma$ is the temporal growth rate, $I_p = (|\hat{E}_{e}|^2 / 16 \pi n_e T_e)$ is the normalized 
plasma wave intensity, and $\omega_e$ is the plasma frequency $\omega_e = \sqrt{4 \pi e^2 n_e / m_e}$ at electron density $n_e$.
In Eq. (\ref{eq5}), $\gamma_{i}$ and $\gamma_p$ are damping rates for the ion-acoustic and electron plasma waves, respectively 
(with plasma wave damping rate equal to the sum of collisonal $\gamma_c$ and Landau $\gamma_L$ damping rates: $\gamma_p = \gamma_c + \gamma_L$).  
The instability threshold determined by Eq. (\ref{eq5}) is readily exceeded during the growth stage of the primary SRS instability 
due to high plasma wave intensity. 

The low-frequency density perturbations generated by the transverse instability seed small-angle SBS 
of the laser light. This manifests in Fig. \ref{fig:f4}(a) as the spreading in $k_y$ about $k_x \approx k_0$. 
The spreading has an equivalent spatial scale of $\sim 5 \lambda_0$ and, through the ion-acoustic sound speed, 
an associated time scale of $\sim 1 \textnormal{ps}$ consistent with Fig. \ref{fig:f3}(a). 

Notably, the spectra also exhibit secondary nonlinear processes that coexist with the dynamic saturation. 
The ion-acoustic and electron plasma wave features around $k_x \approx \pm 2k_0$ and $k_x \approx -k_0$, respectively, 
indicate that SRS-generated electron plasma waves have undergone the Langmuir decay instability. 
The resulting ion-acoustic fluctuations at $k_x \approx 2k_0$ are nearly phase matched to seed Brillouin backscattering 
of the incident light and contribute to an SBS reflectivity of a few percent. Further, the  Raman light spectra show spreading 
in $k_y$---a signature of filamentation. 
As seen in Fig. \ref{fig:f1}(d), the Raman filamentation has a transverse scale $ \gtrsim 10 \lambda_0$, 
but the Raman light is strongly absorbed before propagating out of the instability region.

The {\it LPSE} model for the dynamic saturation of absolute SRS includes
three time-enveloped wave equations for the electric fields of the laser light
${\bf E}_0 = \rm{Re}[\hat{\bf E}_0({\bf r},t)\,\exp(-\mathit{i \omega_0 t})]$,
the Raman scattered light
${\bf E}_1 = \rm{Re}[\hat{\bf E}_1({\bf r},t)\,\exp(-\mathit{i \omega_1 t})]$,
and the electron plasma wave
${\bf E}_e = \rm{Re}[\hat{\bf E}_\mathit{e}({\bf r},t)\,\exp(-\mathit{i \omega_e t})]$,
and a fourth, non-enveloped equation for the low-frequency density perturbation $\delta n$:
\begin{eqnarray}
i {\partial \hat{\bf E}_0 \over \partial t} + i \gamma_0 \circ \hat{\bf E}_0 + 
{c^2 \over 2 \omega_0} \left[ \nabla^2 \hat{\bf E}_0 - 
\nabla (\nabla \cdot \hat{\bf E}_0)  \right]
\nonumber \\
- {\omega_{p}^2 ({\bf r}) - \omega_0^2 \over 2 \omega_0} \hat{\bf E}_0  = 
- {e \over 4 \omega_1 m_e} ( \nabla \cdot \hat{\bf E}_p) \hat{\bf E}_1,
\label{eq1}\\
i {\partial \hat{\bf E}_1 \over \partial t} + i \gamma_1 \circ \hat{\bf E}_1 + 
{c^2 \over 2 \omega_1} \left[  \nabla^2 \hat{\bf E}_1- 
\nabla (\nabla \cdot \hat{\bf E}_1 )  \right]
\nonumber \\
- {\omega_{p}^2 ({\bf r}) - \omega_1^2 \over 2 \omega_1} \hat{\bf E}_1   = 
- {e \over 4 \omega_0 m_e} ( \nabla \cdot \hat{\bf E}_p^*) \hat{\bf E}_1,
\label{eq2}\\
i {\partial \hat{\bf E}_e \over \partial t} + i \gamma_e \circ \hat{\bf E}_e + 
{3 v_{Te}^2 \over 2 \omega_e} \nabla^2 \hat{\bf E}_e  
\nonumber \\
- {\omega_{p}^2 ({\bf r}) - \omega_e^2 \over 2 \omega_p} \hat{\bf E}_e =  
{e \omega_e \over 4 m_e \omega_0 \omega_1  }  \nabla (\hat{\bf E}_0 \cdot \hat{\bf E}_1^* ), 
\label{eq3}\\
{\partial^2 \delta n \over \partial t^2} + 2 \gamma_{i} \circ {\partial \delta n \over \partial t} - 
c_s^2 \nabla^2 \delta n  = 
\nonumber \\
= {1 \over 16 \pi m_i}  \nabla^2 \left[|\hat{\bf E}_e|^2  + {n_e \over n_c}  
\left( |\hat{\bf E}_0|^2 + {\omega_0^2 \over \omega_1^2} |\hat{\bf E}_1|^2  \right) \right], \label{eq4}
\end{eqnarray}
where $\omega_{p} = \sqrt{4 \pi e^2 (n+\delta n) / m_e}$ is the local plasma frequency, $v_{Te} = \sqrt{T_e/m_e}$ is the electron thermal velocity, 
$\circ$ denotes a convolution, and $\gamma_0$, $\gamma_1$, and $\gamma_p$ are the damping rates for each wave. 
The carrier frequencies satisfy the Manley-Rowe relation $\omega_0 = \omega_1 + \omega_e$, with a density $n_e = m_e \omega_e^2 / 4 \pi e^2$ 
selected within the range of the simulation, e.g., close to $n_c/4$. Inverse Bremsstrahlung damping was used for the damping rates of the pump and Raman
light waves. The electron plasma waves were damped by both collisional damping, at a rate of 1 ps$^{-1}$, and Landau damping. The low-frequency perturbations were damped solely by Landau damping at a rate $\gamma_{i} = 0.1 k c_s$ for a fluctuation wave vector $k$. 

By retaining the second-order spatial derivatives in Eqs. (\ref{eq1}) - (\ref{eq4}), {\it LPSE} avoids making the paraxial approximation and can 
model waves propagating in arbitrary directions in inhomogeneous plasmas. The electron plasma waves generated during absolute SRS had $k_e
\lambda_{De} \approx 0.15$, where $k_e \approx (2 \pi / \lambda_0) \sqrt{1 - n_e/n_c})$ and $\lambda_{De} = \sqrt{T_e / 4 \pi e^2 n_e}$, 
indicating that the dynamics occur in a fluid-like regime and that the inclusion of Landau damping 
is sufficient to capture the relevant kinetic effects \cite{Y,MD}.

A novel regime for the saturation of the absolute SRS instability has been shown to mitigate pump depletion 
and increase the transmission of laser light deep into the corona for parameters relevant to ignition-scale direct-drive implosions. 
The dynamic saturation occurs when spatiotemporal fluctuations in the ion-acoustic density 
detune the SRS instability resonance. 
More specifically, the SRS-generated electron plasma waves undergo a transverse instability 
that drives a broad spectum of ion-acoustic fluctuations. 
These fluctuations also seed near-forward Brillouin scattering of the laser light. 
The self-induced spatial incoherence answers, for the first time, a long-standing question in ICF: 
How does laser energy reach the inner corona when an absolute instability region stands in its way.

We acknowlegde useful conversations with M. J. Rosenberg, R. K. Follett, A. A. Solodov, D. P. Turnbull, and D. H. Froula.
This material is based upon work supported by the Department of Energy National Nuclear
Security Administration under Award Number DE-NA0001944, the University of Rochester, and
the New York State Energy Research and Development Authority.

This report was prepared as an account of work sponsored by an agency of the U.S. Government. 
Neither the U.S. Government nor any agency thereof, nor any of their employees, 
makes any warranty, express or implied, or assumes any legal liability or responsibility 
for the accuracy, completeness, or usefulness of any information, apparatus, product, 
or process disclosed, or represents that its use would not infringe privately owned rights. 
Reference herein to any specific commercial product, process, or service by trade name, 
trademark, manufacturer, or otherwise does not necessarily constitute or imply its endorsement, 
recommendation, or favoring by the U.S. Government or any agency thereof. 
The views and opinions of authors expressed herein do not necessarily state or reflect 
those of the U.S. Government or any agency thereof.

\end{document}